\documentclass[sigconf]{acmart}

\AtBeginDocument{%
  \providecommand\BibTeX{{%
    \normalfont B\kern-0.5em{\scshape i\kern-0.25em b}\kern-0.8em\TeX}}}

\copyrightyear{2020}
\acmYear{2020}
\setcopyright{iw3c2w3}

\acmConference[WWW '20]{Proceedings of The Web Conference 2020}{April 20--24, 2020}{Taipei, Taiwan}
\acmBooktitle{Proceedings of The Web Conference 2020 (WWW '20), April 20--24, 2020, Taipei, Taiwan}
\acmPrice{}
\acmDOI{10.1145/3366423.3380016}
\acmISBN{978-1-4503-7023-3/20/04}

\usepackage{graphicx}
\usepackage{tabularx}
\usepackage{amsmath}
\usepackage{bm}
\usepackage{caption}
\usepackage{subcaption}
\usepackage{algorithm}
\usepackage{algorithmic}
\usepackage{graphics} 
\usepackage{multirow}
\usepackage{float}
\usepackage{paralist}
\usepackage{placeins} 
\usepackage{natbib}
\usepackage{mathtools}
\DeclarePairedDelimiter{\ceil}{\lceil}{\rceil}

\newcommand{\mathbbm}[1]{\text{\usefont{U}{bbm}{m}{n}#1}}

\newcommand{\ourcaption}[1]{\caption{\small #1}}

\def\VN{V_{n}}
\def\EN{E_{n}}
\def\NN{N_{n}}
\def\wn{\omega_{n}}
\def\vn{v_{n}}
\def\un{u_{n}}
\def\en{e_{n}}
\def\GN{G_{n}}
\def\ENC{E_{n}(c)}
\def\xnc{x_{n}(c)}
\def\GN{G_{n}}

\newcommand{\separator}{
  \begin{center}
    \rule{\columnwidth}{0.3mm}
  \end{center}
}
\begin{document}

\title{How Much and When Do We Need Higher-order Information in Hypergraphs? A Case Study on Hyperedge Prediction}


\author{Se-eun Yoon, Hyungseok Song, Kijung Shin, and Yung Yi}
\affiliation{%
	\institution{Korea Advanced Institute of Science and Technology}
	\city{Daejeon}
	\country{South Korea}
}
\email{{seeuny,7590sok,kjiungs,yiyung}@kaist.ac.kr}

%
%
%

\renewcommand{\shortauthors}{Yoon et al.}

\begin{abstract}
Hypergraphs provide a natural way of representing group relations, whose complexity motivates an extensive array of prior work to adopt some form of abstraction and simplification of higher-order interactions.
However, the following question has yet to be addressed: 
How much abstraction of group interactions is sufficient in solving a hypergraph task, and how different such results become across datasets? This question, if properly answered, provides a useful engineering guideline on how to trade off between complexity and accuracy of solving a downstream task. 
To this end, we propose a method of incrementally representing group interactions using a notion of {\em $n$-projected graph} whose accumulation contains information on up to $n$-way interactions, and quantify the accuracy of solving a task as $n$ grows for various datasets. As a downstream task, we consider hyperedge prediction, an extension of link prediction, which is a canonical task for evaluating graph models. 
Through experiments on 15 real-world datasets, we draw the following messages: {\bf (a) Diminishing returns:} small $n$ is enough to achieve accuracy comparable with near-perfect approximations, \space {\bf (b) Troubleshooter:} as the task becomes more challenging, larger $n$ brings more benefit, and {\bf (c) Irreducibility:} datasets whose pairwise interactions do not tell much about higher-order interactions lose much accuracy when reduced to pairwise abstractions. 
\end{abstract}



\maketitle

\section{Introduction}

Graphs cover a wide range of applications, but there are domains in which an ordinary graph would fail to capture the relations of entities. Consider a research community, where authors publish papers in groups of more than two. It would involve information loss to represent such groups of collaborators as just pairwise edges as in an ordinary graph. Such interactions are effectively captured by {\em hyperedges,} an extended notion of edges that join an arbitrary number of entities. Graphs with hyperedges, referred to as {\em hypergraphs,} are everywhere in our offline/online networks. People gather in groups \cite{sinha2015MAG}, biological phenomena are caused by joint protein interactions \cite{navlakha2010power}, and web posts contain tags \cite{zhang2019language, ofli2017saki}. 

One of the critical issues in playing with hypergraphs is how to process, simplify, and represent higher-order interactions for a given task. 
One may make a highly abstract representation of complex multi-way interactions, e.g., \cite{zhang2018beyond, yadati2018link, benson2018simplicial, xu2013hyperlink, li2013link}, while others may use the original hypergraph as it is, e.g., \cite{sharma2014predicting, arya2018exploiting, huang2015scalable, yadati2018hypergcn, feng2019hypergraph, benson2018sequences}. 
Despite the recent advances in processing units and memory devices for high-performance data processing, it is still daunting and computationally intractable to maintain whole group interactions in large-scale hypergraphs and use them for solving a given task. 

We are motivated by such a reality and ask the following question: 
How much abstraction of group interactions is sufficient in solving a given graph task, and how different such results become across datasets that vary in scale, entities, and pattern of interactions? The answers to this question would give us useful engineering guidelines on how to appropriately trade off between complexity in representation of higher-order group interactions and accuracy of solving the task. In seeking to answer this question, we may find a new method that outperforms existing algorithms in literature while maintaining computational tractability. 
In this paper, we consider the hyperedge prediction task, which is a hypergraph extension of link prediction. 
Link prediction is a widely accepted means of assessing the validity of graph models \cite{liben2007link,lu2011link, grover2016node2vec,zhou2017scalable,santolini2018predicting,you2019position,grover2019graphite}.

As an important device to answer our question by solving the hyperedge prediction task, we introduce the {\em $n$-projected graph}, $\GN$, $n=2, 3, \ldots$ for a given hypergraph $G$. 
This is a modified version of the original hypergraph $G$ so as to contain $n$-way group interactions.
By incrementally stacking $n$-projected graphs, we can represent the original hypergraph with up to $n$-way interactions. As expected, as $n$ grows, we reduce the information loss, but the computational cost for processing increases. 
The notion of projected graphs is not entirely new as adopted in \cite{zhang2018beyond, yadati2018link, benson2018simplicial, sharma2014predicting}. However, it has been limited to the \textit{pairwise} relation, which turns out to be the $2$-projected graph, a special case of the $n$-projected graph. 
We generalize this pairwise relation to $n$-way interactions in $\GN$ to quantify and decompose the degree of interactions, constructed as follows: 
Each edge in $\GN$ is weighted by the number of hyperedges in which the node set of size $n$ have appeared together (see Section~\ref{sec:methods} for details). 

The value of the $n$-projected graph is clear by the following example: Suppose that  
we want to predict whether four people would collaborate or not in the future. 
It is useful to know how much each pair has collaborated together as in $2$-projected graph. However, collaboration is often formed because a group of three people, who have collaborated as a group, may recruit a fourth person in the future, where $3$-way interaction becomes valuable.

We conduct experiments using 15 datasets spanning 8 domains provided by \cite{benson2018simplicial, stehle2011high, mastrandrea2015contact, yin2017local, leskovec2007evolution, fowler2006connecting, fowler2006cosponsorship, sinha2015MAG}. 
These datasets are highly heterogeneous in terms of scale, pattern of interactions, and interacting entities, ranging from about 1,000 to 2,500,000 hyperedges. 
We use logistic regression for prediction, where we utilize the features popularly used in link/hyperedge prediction tasks but generalized for $n$-projected graphs. The prediction results would change for different methods, but we experience similar trends. 
We summarize the key findings of our experiments in what follows: 

\smallskip
\begin{compactitem}[$\circ$]
\item {\bf Diminishing returns.} We systematically analyze the gain of approximating a hypergraph with increasing orders of $n$. Particularly, we find that small orders of $n$ are enough to achieve comparable accuracy with near perfect approximations.

\smallskip
\item {\bf Troubleshooter.} As we explore the outcomes in possible variations of the task, we discover that higher-order helps more in more challenging variations. 

\smallskip
\item {\bf Irreducibility.} We search for theoretical interpretations as to why the benefit of higher $n$ is greater in some datasets than in others. These are datasets whose higher-order relations share little information with pairwise relations, thus cannot be reduced to pairwise.
\end{compactitem}

Our source code and appendix are available online at \cite{appendix}.

\section{Related work}

Hypergraphs have been used in various domains, including social networks \cite{tan2014mapping, yang2019revisiting}, text retrieval \cite{hu2008hypergraph}, recommendation \cite{bu2010music, zhu2016heterogeneous}, knowledge graphs \cite{fatemi2019knowledge}, bioinformatics \cite{klamt2009hypergraphs, hwang2008learning}, e-commerce \cite{li2018tail}, computer vision \cite{huang2015learning, chen2009efficient} and circuit design \cite{ouyang2002multilevel, karypis1999multilevel}. 
Learning tasks based on hypergraphs include clustering \cite{zhou2007learning, agarwal2005beyond, karypis2000multilevel, huang2015scalable}, classification \cite{yadati2018hypergcn, feng2019hypergraph}, and hyperedge prediction \cite{zhang2018beyond, yadati2018link, benson2018simplicial, xu2013hyperlink, li2013link, sharma2014predicting, arya2018exploiting, benson2018sequences}.

\smallskip
\noindent\textbf{Hypergraph representation.} 
To represent hypergraphs in an abstract manner, one method is to perform dyadic projection, also known as the clique expansion, reflecting two-way node relationships. This leads to usage of powerful tools such as spectral clustering \cite{zhou2007learning}. Clique averaging is a similar method \cite{agarwal2005beyond} which assigns edge weights differently. \citet{karypis2000multilevel} create successively coarser versions of a hypergraph for partitioning. 
The category of using hypergraphs without modification includes star expansion \cite{agarwal2006higher} that connects each node in a hyperedge to a new node that represents a hyperedge. There are works that directly use hypergraphs with the idea of two resilient distributed datasets (RDDs) \cite{huang2015scalable} and deep learning approaches \cite{yadati2018hypergcn, feng2019hypergraph}.

\smallskip
\noindent\textbf{Representation in hyperedge prediction.}
We now focus on prior works on hyperedge prediction. 
There are works that handle hypergraphs just with pairwise relations. \citet{zhang2018beyond} project a hypergraph into a dyadic graph and uses its adjacency matrix for factorization. \citet{yadati2018link} propose a deep learning approach with a 2-projected graph as the input. \citet{benson2018simplicial} compare the performances of various features from the 2-projected graph to predict the co-occurrence of node triples. \citet{xu2013hyperlink} learn representations for the distance matrix constructed from dyadic hops. \citet{li2013link} rank hyperedges according to the proximity between two users. 
Another array of research apply hypergraphs as they are, implying the importance of using higher-order interactions. \citet{sharma2014predicting} claim that 2-projected graphs fail to capture higher-order relationships. \citet{arya2018exploiting} represent the whole hypergraph as the matrix of a star-expanded graph \cite{agarwal2006higher} and formulate hyperedge prediction as a matrix completion problem. \citet{benson2018sequences} operate on the sequence of sets, a timestamped representation of hyperedges, to generate the next timestamp hyperedge. 
 
In this paper, we propose a parameterized representation framework that generates the entire spectrum of projected graphs and study the impact of the degree of simplification. An additional benefit of $n$-way decomposition as in our $n$-projected graphs is that each degree allows a certain form of uniformity, which enables us to enjoy computational amenity and mathematical tractability at each $n$. Such benefits are verified in other contexts by \cite{kolda2009tensor,shashua2006multi, bulo2009game, ghoshdastidar2017uniform,lin2009metafac}.

\section{Problem formulation}
\label{sec:problem}
In this section, we formulate the problem of hyperedge prediction (Sections~\ref{subsec:hypergraphs} and \ref{subsec:hyperedge_prediction}), which serves as a tool to evaluate the accuracy of hypergraph abstractions, and introduce possible variations on the problem (Section~\ref{subsec:candidate}).

\subsection{Concepts: Hypergraphs}\label{subsec:hypergraphs}

Let $G=(V,E,\omega)$ be a hypergraph where $V$ is a set of nodes and $E\subseteq 2^{V}$ is a set of hyperedges. Each hyperedge $e\in E$ represents a set of $|e|$ nodes that took interaction. 
We weight each hyperedge by the number of times occurrence, and each hyperedge $e$ has a positive weight $\omega(e)$.

\subsection{Problem: Hyperedge prediction}\label{subsec:hyperedge_prediction}

The problem of hyperedge prediction is generally defined as: Given an hypergraph in which hyperedges have timestamps up to $t$, predict the hyperedges that will appear from $t$ until a time point $t' > t$ in the future. However, a common practice is to remove some hyperedges from a snapshot of a hypergraph and regard them as the ones in the future \cite{grover2016node2vec, yadati2018link}, since timestamps are unavailable in many real-world data.\footnote{Though the datasets in this paper are originally timestamped, we follow this practice.} Furthermore, it is unnecessary to generate all the missing hyperedges from $\mathcal{O}(2^{|V|})$, since the extreme sparsity would lead to poor generalization \cite{zhang2018beyond}. 
Thus, we solve a standard binary classification problem (Problem~1), where we use $E'$ to indicate the set of hyperedges remaining after some are removed from $E$:

\vspace{-0.15cm}
\separator
\vspace{-0.1cm}
\noindent\textbf{Problem 1} \textsc{(Hyperedge prediction).}

\begin{compactitem}[$\circ$]
\item {\bf Given:} 
\begin{itemize}
\item a hypergraph $G=(V, E')$

\vspace{0.05cm}
\item a candidate hyperedge set $C \subseteq 2^{V}$ where $C\cap E' = \emptyset$ 

\end{itemize}
\vspace{0.1cm}
 \item {\bf Decide:} whether each subset $c\in C$ belongs to $E$ where $E'\subseteq E$.

\end{compactitem}
\vspace{-0.3cm}
\separator

We divide $C$ into a set $C_p$ of positive hyperedges in $E$ and a set $C_n$ of negative hyperedges not in $E$.
That is, $C_p \subseteq E$, while $C_n \cap E= \emptyset$.
Then, the objective is to find a classifier $f: C \rightarrow \{0, 1\}$ that is close to the perfect classifier $f^{\star}$, where  $f^{\star}(c)=1$ for $c \in C_p$ and $f^{\star}(c)=0$ for $c \in C_n$.

\subsection{Constructing hyperedge candidate set $C$}\label{subsec:candidate}  

There are different ways of constructing the candidate set $C$. We thoroughly examine different choices of $C$ since experiments on a single choice could be biased for that particular case.

\noindent\textbf{Hyperedge size.} 
We consider three cases where each candidate $c$ has cardinality $4$, $5$, and $10$, respectively.
For each size, we systematically analyze the effect of higher-order interactions.

\noindent\textbf{Negative hyperedges.} While positive hyperedges $C_p$ can be collected simply by removing a certain proportion of $E$, negative hyperedges $C_n$ need to be generated from $2^V \setminus E.$ If the nodes in each negative hyperedge are independently sampled, the resulting hyperedge will be unlikely to occur (e.g., total strangers are very unlikely to collaborate), making classification trivial. To avoid this situation, we select nodes that form \textit{stars} or \textit{cliques} in the pairwise projected graph as negative hyperedges. From the pairwise perspective, nodes that form a clique are more strongly tied and thus more likely to form a hyperedge than those which form a star. Thus, the task becomes more challenging when $C_n$ is generated from cliques.

\noindent\textbf{Class imbalance.} Now that we have considered the quality of negative hyperedges, we turn our attention to their quantity: how large should $C_n$ be? Since only a few form hyperedges among all possible node sets, it is natural to make $|C_n| \geq |C_p|$, imposing class imbalance. We set the class ratio $|C_p|:|C_n|$ to be 1:1, 1:2, 1:5, 1:10, and for some cases, 1:200. Larger imbalance adds more difficulty to finding all $C_p$ while being precise as not to falsely predict $C_n$. 
\section{Methods}\label{sec:methods}
In this section, we formally define the $n$-projected graph and the $n$-order expansion (Section~\ref{subsec:pg}), and we describe our prediction model based upon the $n$-order expansion (Section~\ref{subsec:model}).

\subsection{The n-order expansion}\label{subsec:pg}
We propose a method of incrementally representing high-order interactions in a given hypergraph, namely the $n$-order expansion. Each increment in the representation is given as the $n$-projected graph (or $n$-pg in short), which captures the interactions of $n$ nodes. We note that there could be other ways of extracting uniform-size interactions, but we choose the $n$-pg since its graphical representation enables the adoption of various principled link prediction features that are widely acknowledged in literature \cite{adamic2003friends,benson2018simplicial, liben2007link, grover2016node2vec}. Furthermore, it is a generalization of the commonly-used pairwise projected graph, providing conceptual consistency. 

\begin{definition}[$n$-projected graph]
The $\mathbf{n}$\textbf{-projected graph} $\GN=(\VN, \EN, \wn)$ of a hypergraph $G=(V, E, \omega)$ is defined as follows:
\begin{align*}
	 & \VN  := \{\vn \subseteq V  : |\vn| = n -1\}, \\
	 & \EN  := \{(\un,\vn) \in \VN^2 :  |\un \cup \vn| = n \text{ and } \ \exists e\in E \text{ s.t. } \un \cup \vn \subseteq e \}, \\
	 & \wn \big((\un,\vn) \big) :=\sum\nolimits_{e \in E} \omega(e) \cdot \mathbbm{1}\big[(\un \cup \vn) \subseteq e \big].
\end{align*}
\end{definition}

\begin{figure}[!t]
  \includegraphics[width=0.99\linewidth]{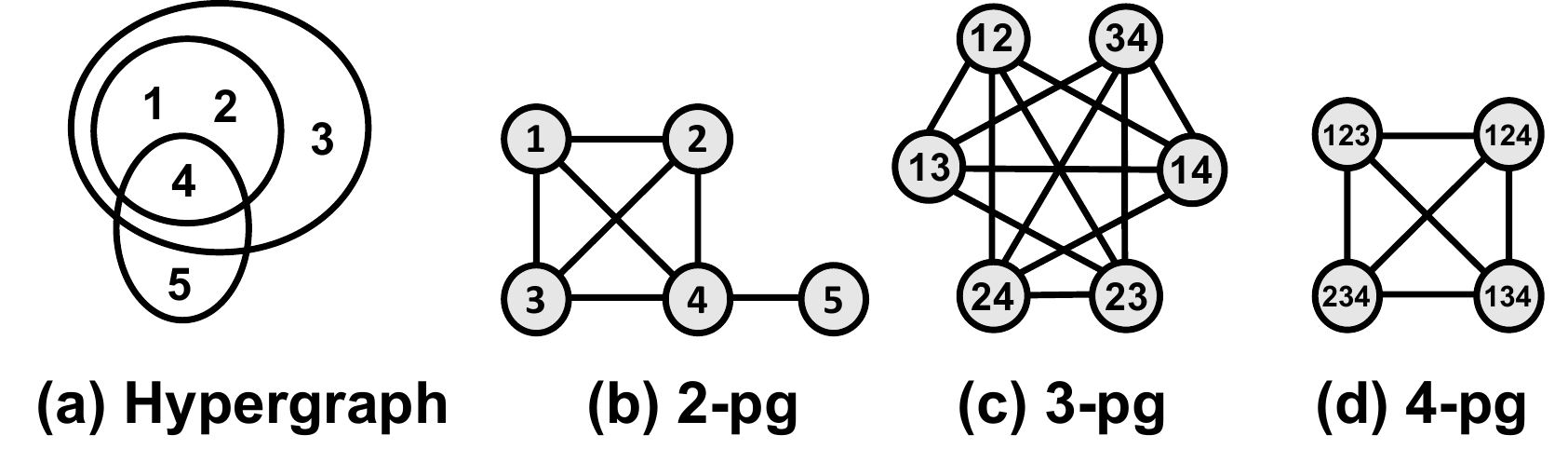}
  \ourcaption{Hypergraph and its $n$-projected graphs ($n$-pgs). Edge weights, dropped for simplicity, reflect $n$-way interactions. For example, edge weight between the 3-pg nodes \{1, 2\} and \{2, 3\} is the number of times \{1, 2, 3\} interacted as a group.
}
\label{fig:concept}
\end{figure}

\noindent
That is, each node $\vn$ in the $n$-projected graph $\GN$ of a hypergraph $G$ is a size-($n$-$1$) subset of nodes in $G$, and each edge $(\un,\vn)$ represents a size-$n$ subset of nodes in $G$ contained in at least one hyperedge in $G$. 
The weight of an edge corresponds to the sum of weights of hyperedges in $G$ that contain the size-$n$ subset represented by the edge.
In other words, the weight of each edge indicates how often the corresponding $n$ nodes interact as a group and thus how close they are as a group.
Notice that the pairwise projected graph is a special case of the $n$-projected graph where $n=2$. Figure~\ref{fig:concept} gives a visual description.

Based on $n$-projected graphs, we define the $n$-order expansion, our proposed way of incrementally approximating a hypergraph.
\begin{definition}[$n$-order expansion]
The $\mathbf{n}$\textbf{-order expansion} $G^{(n)}$ of a hypergraph $G$ is a collection of $k$-projected graphs where $k$ varies from $2$ to $n$. That is, 
$G^{(n)} := (G_{2}, \ldots, \GN).$
\end{definition}

As $n$ increases, the $n$-order representation $G^{(n)}$ captures more information in $G$, and if $n$ reaches its maximum, $G$ can be reconstructed from $G^{(n)}$.
In Section~\ref{sec:exp}, we experimentally study the value of marginal information gain (quantified by prediction accuracy) for each $n$ in the $n$-order expansion in hyperedge prediction. 

\subsection{Prediction model}\label{subsec:model}
In this subsection, we describe the features and classifier that we use for hyperedge prediction.

\smallskip
\noindent\textbf{Features.} 
The $n$-order expansion of a hypergraph $G$ returns a series of $n$-projected graphs, from each of which we extract one among six features.
Let $\NN(\vn)$ be the set of neighbors of the node $\vn$ in the $n$-projected graph $\GN$, and for each subset $c$ of nodes in $G$, let $\ENC:=\{(\un,\vn)\in \EN: \un \subseteq c \text{ and } \vn\subseteq c \}$ be the set of ``inner'' edges in $\GN$ that represent a subset of $c$. 
Then, we use the following features extractable from $\GN$ for each hyperedge candidate $c \in C$:

{\def\arraystretch{2}
\resizebox{0.97\linewidth}{!}{%
\hspace{-6mm}
\begin{tabular}{ll}
    $\circ$ \textbf{Geometric mean (GM):} & 
    $\xnc=\Big(\prod_{\en \in \ENC} \omega_n(\en)\Big)^\frac{1}{|\ENC|} $ \\
    $\circ$ \textbf{Harmonic mean (HM):} & 
    $\xnc=\frac{|\ENC|}{\sum_{\en \in \ENC} \omega_{n}(e_{n})^{-1}}$ \\
    $\circ$ \textbf{Arithmetic mean (AM):} & 
    $\xnc=\frac{1}{|\ENC|}\sum_{\en \in \ENC} \omega_{n}(e_{n})$ \\
    $\circ$ \textbf{Common neighbors (CN):} &
    $\xnc= \bigcap_{\vn \subseteq c} N_n(\vn)$ \\
    $\circ$ \textbf{Jaccard coefficient (JC):} & 
    $\xnc= \frac{\bigcap_{\vn \subseteq c} N_n(\vn)}{\bigcup_{\vn \subseteq c} N_n(\vn)}$ \\
    $\circ$ \textbf{Adamic-Adar index (AA):} &
    $\xnc= \sum\limits_{\un \in \bigcap_{\vn\subseteq c} N_n(\vn)} \frac{1}{\log|N_n(\un)|}$ \\
\end{tabular}}
}

The first three measures (GM, HM, and AM) are the geometric, harmonic, and arithmetic means of inner edge weights in the $n$-projected graphs.  
These features are reported to work well in the task of predicting triangles in $2$-projected graphs \cite{benson2018simplicial}. 
For the other three measures (CN, JC, and AA), we extend well-known pairwise link prediction features \cite{newman2001clustering, salton1983introduction,adamic2003friends} to larger groups of nodes.

When the input hypergraph is represented in the form of the $n$-order expansion $G^{(n)}$, the features obtained in different projected graphs are concatenated.
That is, the features of a subset $c$ of nodes obtained from $G^{(n)}$ are $[x_{2}(c), \ldots, \xnc]$. 

\smallskip
\noindent\textbf{Classifier.} 
We use the above features as the inputs to a logistic regression classifier with L2 regularization, which has been used widely for link and hyperedge prediction \cite{grover2016node2vec, benson2018simplicial, liben2007link}. 
Although complicated classifiers with more parameters, such as deep neural networks, could be used instead, their performance has higher variance and depends more heavily on hyperparameter values.
We decide to use the simple classifier to provide stable comparisons of different orders of approximation.

\section{Experiments}
\label{sec:exp}

In this section, we present our experimental results to address our questions on the impact of higher-order interactions in the form of $n$-projected graphs (or simply $n$-pgs throughout this section). 

\subsection{Setup}

We start by explaining our datasets and the experimental setup, followed by our results in each of subsections. 

\smallskip
\noindent\textbf{Datasets.} We use 15 datasets generated across 8 domains from \cite{benson2018simplicial}\footnote{\url{https://www.cs.cornell.edu/~arb/data/}}. The numbers of edges and hyperedges in them are summarized in Table~\ref{tab:datasets}. 
Hyperedges in each domain are defined as follows:
    {\bf (a) Email} (email-Enron \cite{klimt2004enron}, email-Eu \cite{yin2017local}): recipient addresses of an email,
    {\bf (b) Contact} (contact-primary-school \cite{stehle2011high}, contact-high-school \cite{mastrandrea2015contact}): persons that appeared in face-to-face proximity,
    {\bf (c) Drug components} (NDC-classes, NDC-substances): classes or substances within a single drug, listed in the National Drug Code Directory,
    {\bf (d) Drug use} (DAWN): drugs used by a patient, reported to the Drug Abuse Warning Network, before an emergency visit,
    {\bf (e) US Congress} (congress-bills \cite{fowler2006cosponsorship}): congress members cosponsoring a bill,
    {\bf (f) Online tags} (tags-ask-ubuntu, tags-math-sx): tags in a question in Stack Exchange forums,
    {\bf (g) Online threads} (tags-ask-ubuntu, tags-math-sx): users answering a question in Stack Exchange forums, and
    {\bf (h) Coauthorship} (coauth-MAG-History \cite{sinha2015MAG}, coauth-MAG-Geology \cite{sinha2015MAG}, coauth-DBLP): coauthors of a publication.
We only consider hyperedges containing at most 10 nodes. It is reported that large hyperedges are rare and less meaningful \cite{benson2018simplicial}. As mentioned in Section~\ref{subsec:hyperedge_prediction}, the datasets are timestamped but we treat them as weighted hypergraphs with unique hyperedges. 

\smallskip
\noindent\textbf{Training and evaluation.} For each target hyperedge size (4, 5, 10), we generate positive hyperedges $C_p$ by randomly removing hyperedges until $60\%$ of all hyperedges or no hyperedges with the target size are left. 
We randomly sample the sets of nodes that form cliques and stars in 2-pg as negative hyperedges $C_n$ until a certain multiple of positive hyperedges ($\times$ 1, 2, 5, 10, 200) are gathered (see Section~\ref{subsec:candidate} for details). The positive and negative hyperedges are combined to form the candidate set, i.e., $C=C_p \sqcup C_n$. The candidate set $C$ is split into train $(50\%)$ and test $(50\%)$ sets. We evaluate classification performance by the area under the precision-recall curve (AUC-PR) \cite{davis2006relationship}, a measure sensitive to class imbalance.

\begin{table}[tb]
\vspace{-2mm}
\ourcaption{Dataset statistics. We count the number of unique hyperedges and the number of edges in the 2,3,4-projected graphs. More details are provided in \cite{appendix}.}
\label{tab:datasets}
  \resizebox{\linewidth}{!}{%
  \begin{tabular}{l|c|c|c|c}
    \toprule
    Dataset & $|E|$ & $|E_{2}|$ & $|E_{3}|$ & $|E_4|$ \\
    \hline 
    email-Enron & 1,491 & 1,442 & 8,916 & 25,938 \\
    email-Eu & 24,223 & 21,465 & 143,238 & 440,916 \\
    contact-primary-school & 12,704 & 8,317 & 15,417 & 2,286 \\
    contact-high-school & 7,818 & 5,818 & 7,110 & 1,428 \\
    NDC-classes & 901 & 3,727 & 21,885 & 61,176 \\
    NDC-substances & 8,167 & 26,973 & 234,240 & 729,012 \\
    DAWN & 137,417 & 97,046 & 1,456,683 & 4,917,996 \\
    congress-bills & 57,887 & 178,647 & 2,439,960 & 8,117,514 \\
    tags-ask-ubuntu & 147,222 & 132,703 & 838,107 & 874,056 \\
    tags-math-sx & 170,476 & 91,685 & 748,644 & 936,774 \\
    threads-ask-ubuntu & 166,995 & 186,955 & 181,881 & 116,046 \\
    threads-math-sx & 595,648 & 1,083,531 & 2,184,567 & 2,174,994 \\
    coauth-MAG-History & 891,296 & 723,382 & 2,101,608 & 4,226,058 \\
    coauth-MAG-Geology & 1,189,770 & 4,241,817 & 18,870,564 & 40,067,280 \\
    coauth-DBLP & 2,454,734 & 7,123,888 & 26,398,201 & 46,071,251 \\
  \bottomrule
  \end{tabular}
  }
\end{table}

\begin{table*}[htb]
\vspace{-2mm}
\ourcaption{Performance gain of $n$ to $(n+1)$-order expansion. Gain is computed as percentage improvement in AUC-PR. For predicting size 4 hyperedges, we use 2 and 3-order expansions. For predicting size 5, we use 2, 3, and 4-order expansions. Increasing the order of expansion gives better performance (marked in bold) in most cases, across datasets and features. Results on every task variation are in \cite{appendix}.}
\label{tab:performance}
  \small\addtolength{\tabcolsep}{-3pt}
  \resizebox{\textwidth}{!}{
  \begin{tabular}{lcccccccccccccccccc}
    \toprule
    \multirow{2}{*}{}
    & \multicolumn{6}{c}{\bf Size 4 prediction}
    & \multicolumn{12}{c}{\bf Size 5 prediction}\\
    \cmidrule(lr){2-7}
    \cmidrule(l){8-19}
    & \multicolumn{6}{c}{\bf 2 to 3 gain (\%)}
    & \multicolumn{6}{c}{\bf 2 to 3 gain (\%)}
    & \multicolumn{6}{c}{\bf 3 to 4 gain (\%)}\\
    \cmidrule(lr){2-7}
    \cmidrule(lr){8-13}
    \cmidrule(l){14-19}
    \textbf{Dataset}
    & GM & HM & AM & CN & JC & AA
    & GM & HM & AM & CN & JC & AA
    & GM & HM & AM & CN & JC & AA\\
    \midrule
    email-Enron 
    & \bf12.67 & \bf0.49 & -0.15 & \bf33.89 & -1.57 & \bf35.78 
    & \bf14.37& \bf22.59 & \bf5.95 & \bf10.73 & -2.98 & \bf17.01 
    & -1.53 & -0.59 & -0.26 & \bf5.04 & \bf1.59 & -2.05 \\
    email-Eu
    & \bf0.69 & -0.13 & \bf1.67 & \bf153.81 & \bf9.16 & \bf148.09
    & -2.44 & -0.78 & \bf4.04 & \bf158.79 & \bf13.93 & \bf157.73
    & \bf0.64 & -0.32 & \bf1.23 & \bf2.01 & \bf18.86 & \bf2.31 \\
    contact-primary-school 
    & \bf6.42 & \bf1.21 & \bf49.2 & \bf495.94 & \bf413.35 & \bf484.73 
    & 0 & 0 & \bf6.63 & \bf708.56 & \bf361.79 & \bf267.66 
    & 0 & 0 & 0 & 0 & 0 & 0 \\
    contact-high-school
    & \bf15.16 & -0.87 & \bf78.62 & \bf515.17 & \bf455.54 & \bf507.13 
    & 0 & 0 & \bf14.14 & \bf1623.33 & \bf221.51 & \bf1617.75 
    & 0 & \bf5.96 & \bf3.9 & 0 & \bf104.37 & 0 \\
    NDC-classes 
    & \bf0.18 & \bf4.23 & -44.70 & \bf1.55 & \bf10.91 & \bf2.25 
    & \bf44.28 & \bf16.62 & -37.82 & \bf0.28 & \bf10.07 & \bf5.60 
    & \bf6.77 & \bf16.95 & -2.31 & -4.14 & \bf2.00 & -1.50 \\
    NDC-substances 
    & -4.95 & -0.02 & \bf0.47 & \bf0.57 & -40.98 & -3.54
    & \bf10.73 & \bf2.46 & \bf0.16 & \bf16.01 & -17.02 & \bf14.18
    & \bf158.10	& \bf0.02 & \bf7.07 & -1.81 & -0.47 &-2.99 \\
    DAWN
    & \bf0.15 & \bf0.04 & \bf21.34 & \bf197.97 & \bf30.48 & \bf187.62
    & \bf0.23 & \bf3e-4 & \bf3.48 & \bf220.79 & \bf42.80 & \bf212.33 
    & \bf0.49 & \bf4e-4 & \bf14.04 & -0.85 & \bf17.31 & -0.54\\
    congress-bills
    & \bf7.92 & -0.99 & \bf14.53 & \bf328.76 & \bf16.49 & \bf294.16
    & \bf11.84 & -0.03 & \bf30.86 & \bf271.64 & \bf48.55 & \bf259.22
    &-0.07 & \bf4.98 & \bf0.26 & \bf0.93 & \bf0.57 & \bf0.16\\
    tags-ask-ubuntu
    & \bf0.24 & -0.51 & \bf23.09 & \bf216.47 & \bf14.07 & \bf192.03
    & \bf0.07 & \bf0.02 & \bf20.84 & \bf244.72 & \bf80.37 & \bf225.89
    & \bf1e-05 & -1.13 & \bf2.96 & \bf0.85 & \bf5.50 & \bf1.35\\
    tags-math-sx
    & \bf0.46 & \bf0.18 & \bf32.38 & \bf137.4 & \bf46.53 & \bf127.25
    & \bf0.13 & \bf0.01 & \bf21.35 & \bf146.02 & \bf60.64 & \bf135.54
    & \bf1e-05 & \bf0.63 & \bf9.73 & \bf0.67 & \bf5.86 & \bf0.74\\
    threads-ask-ubuntu 
    & \bf2e-3 & \bf10.05 & \bf2.47 & \bf2.34 & \bf2.34 & \bf1.48
    & -1e-3 & -1.76 & \bf6.51 & \bf2.56 & \bf3.10 & \bf1.76
    & -1e-4 & \bf9.62 & -0.07 & \bf0.01 & -0.05 & \bf1e-3 \\
    threads-math-sx
    & \bf0.03 & \bf0.44 & \bf8.52 & \bf6.01 & \bf5.61 & \bf5.10
    & \bf5e-3 & \bf0.42 & \bf23.48 & \bf6.63 & \bf6.56 & \bf5.65
    & \bf1e-3 & \bf0.61 & \bf0.01 & \bf2e-4 & -0.15 & -4e-6\\
    coauth-MAG-History
    & \bf4e-3 & -8.48 & -1.81 & \bf1.69 & \bf3.00 & \bf1.94
    & \bf0.08 & \bf0.13& \bf3.00 & \bf2.32 & \bf4.43 & \bf2.37
    & \bf0.16 & \bf137.94 & \bf2.36 & -0.23 & \bf0.53 & -0.10\\
    coauth-MAG-Geology
    & \bf0.93 & \bf0.36 & \bf22.93 & \bf15.39 & \bf19.76 & \bf15.34
    & \bf0.79 & \bf3.78 & \bf603.02 & \bf14.56 & \bf20.52 & \bf14.65
    & -30.08 & \bf0.17 & \bf8.14 & -0.10 & \bf1.68 & \bf0.39\\
    coauth-DBLP
    & -53.43 & \bf0.90 & \bf145.31 & \bf16.89 & \bf21.99 & \bf16.73
    & \bf1.32 & \bf3.52 & \bf175.24 & \bf16.17 & \bf22.82 & \bf15.44
    &-24.05 & \bf0.58 & \bf11.06 & -0.08 & \bf1.69 & \bf0.30\\
    \cmidrule(r){1-1}
    \cmidrule(lr){2-7}
    \cmidrule(lr){8-13}
    \cmidrule(l){14-19}
    \textbf{Average} & -0.90 & \bf0.46 & \bf23.60 & \bf141.59 & \bf67.11 & \bf134.41 & \bf5.43 & \bf3.16 & \bf58.73 & \bf229.54 & \bf58.47 & \bf220.85 & \bf7.36 & \bf11.69 & \bf3.88 & \bf0.15 & \bf10.62 & -0.13\\
  \bottomrule
  \end{tabular}}
\end{table*}

\subsection{Results and messages}
In this subsection, we present our results by summarizing them with three main messages. 

\vspace{-0.15cm}
\separator
\vspace{-0.15cm}
\noindent{\bf (M1) More higher-order information leads to better prediction quality, but with diminishing returns.}
\vspace{-0.3cm}
\separator
\vspace{-0.1cm}

\begin{figure}[!t]
\vspace{-2mm}
\centering
\begin{subfigure}{\linewidth}
  \includegraphics[width=1\linewidth]{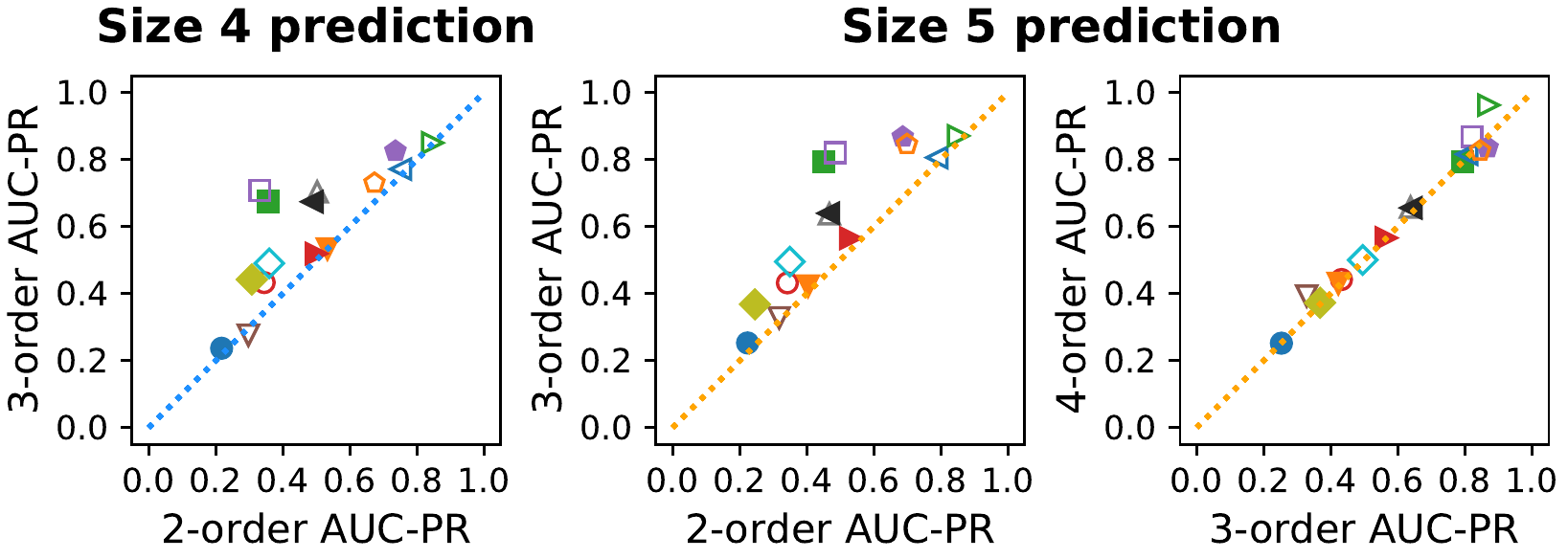}
  \label{fig:clique} 
\end{subfigure}\\ \vspace*{-0.3cm}
\begin{subfigure}{\linewidth}
  \includegraphics[width=1\linewidth]{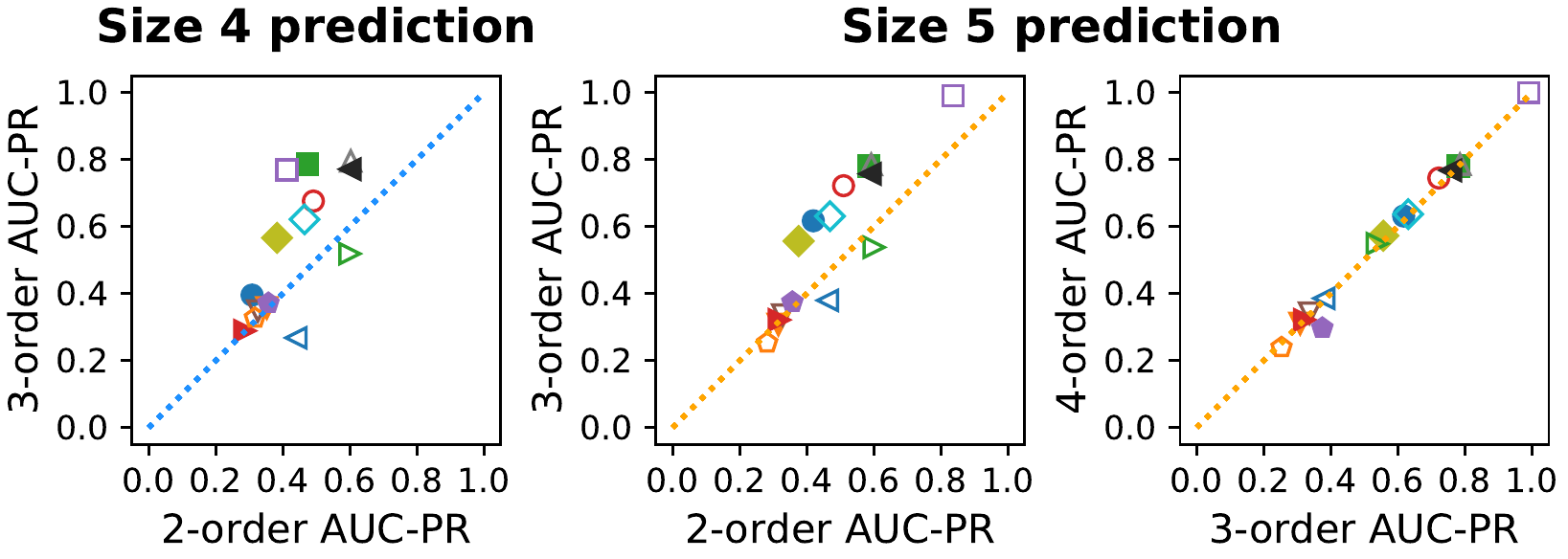}
  \label{fig:star}
\end{subfigure}\\ \vspace*{-0.3cm}
\begin{subfigure}{\linewidth}
  \includegraphics[width=1\linewidth]{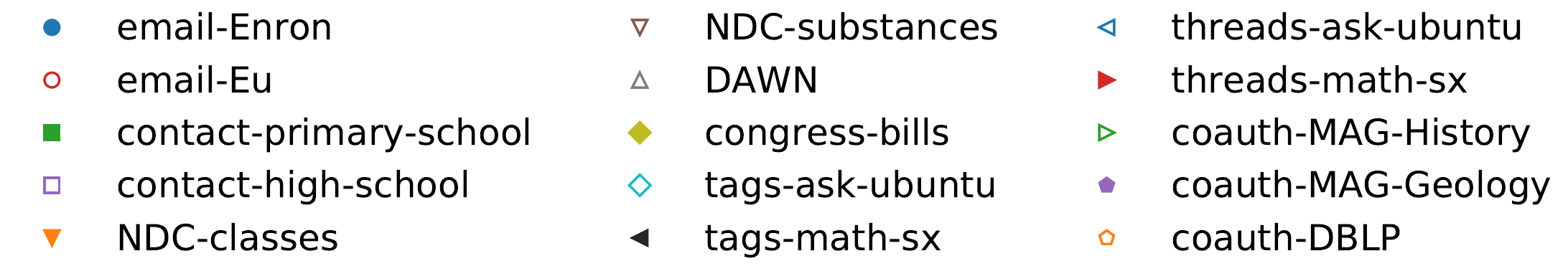}
  \label{fig:legend}
\end{subfigure} \\ \vspace*{-0.3cm}
\ourcaption{Performance comparison between order $n$ and $n+1$. Performance is averaged across features. Nodes that form cliques (upper) and stars (lower) in 2-pg are used as negative hyperedges. Class imbalance is set to 1:10.  While there is overall performance gain, the gain from 3 to 4 is smaller than 2 to 3.}
\label{fig:performance}
\end{figure}

We investigate how the prediction performance changes with increasing $n$ in $n$-order expansions. In particular, we predict hyperedges of size 4 with the features from 2 and 3-order expansions, and hyperedges of size 5 with features from 2, 3, and 4-order expansions. Table~\ref{tab:performance} summarizes the results, and for readers' convenience, we also plot the performance averaged across all features in Figure~\ref{fig:performance}. 
\begin{figure}[t]
    \vspace{-2mm}
  \centering
  \includegraphics[width=\linewidth]{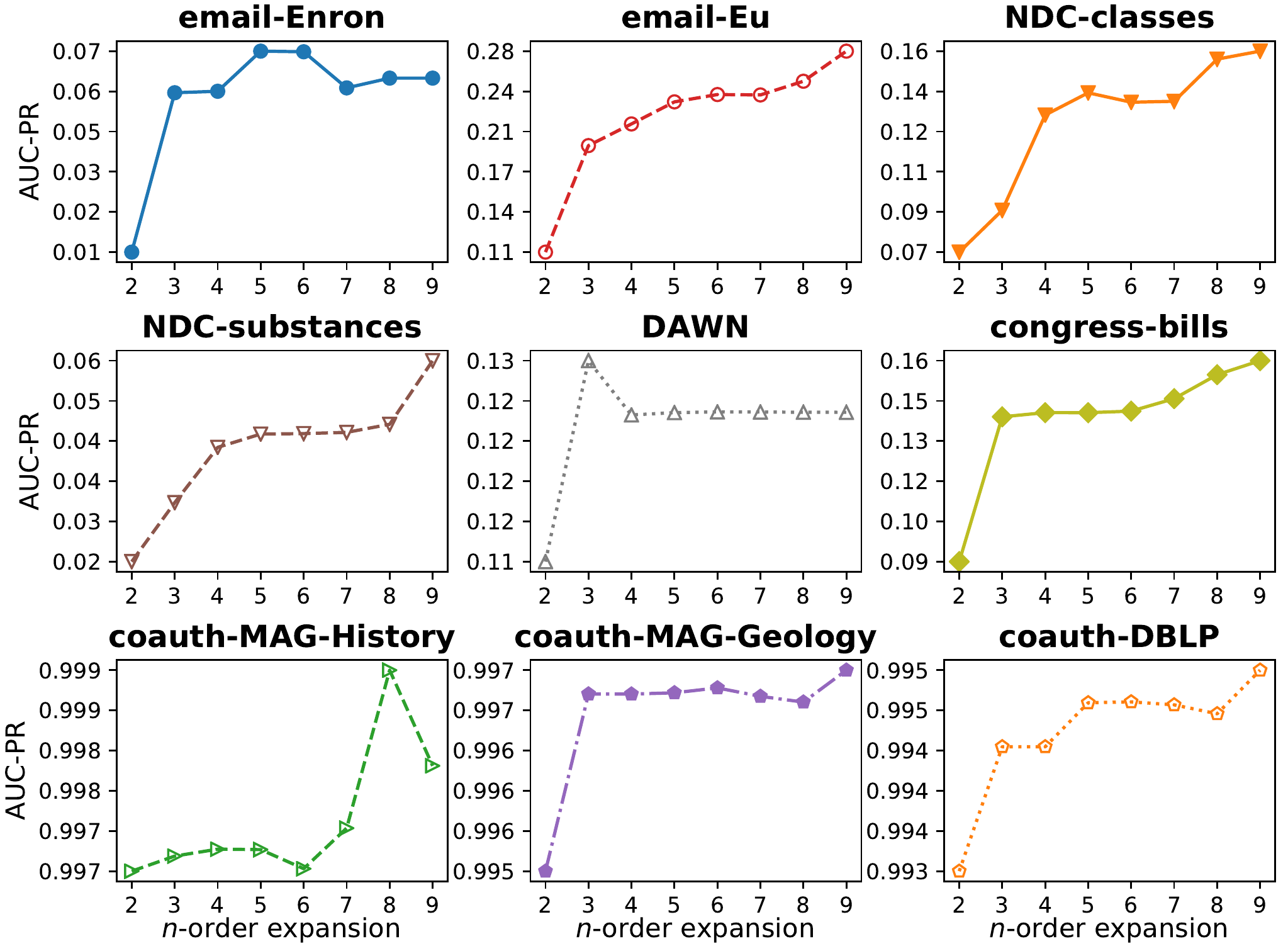}
  \ourcaption{Diminishing returns of $n$-order expansions. We set target size as 10 and observe until 9-order expansions. We exclude datasets that has no edges in any of the projected graphs, and we set the class imbalance to 1:200.
  We take the average performance across all features except the geometric mean, whose value can become too large in higher-order pgs. As $n$ grows, performance tend to increase. However, gains tend to decrease.}
\label{fig:marginal}
\end{figure} 

We clearly observe that higher-order expansion gives better performance, where the improvement quantity differs across datasets and features. Performance gaps from $n=2$ to $n=3$, averaged across features, are 61\% for size 4 and 94\% for size 5, respectively, whereas it is just 6\% for size 5 from $n=3$ to $n=4.$  
As for the individual features, we see that the gain is larger with neighborhood-based features (CN, JC, AA) than with mean-based features (GM, HM, AM). 
The mean values are small in higher-order pgs, while neighborhood-based features still retain meaningfully large values. Entries with exactly zero gain (contact datasets) result from the sparsity of size 5 hyperedges (i.e., $< 10$). See more details in \cite{appendix}. 

Interestingly, we see the diminishing returns as $n$ increases. To study this,   
we predict size 10 hyperedges with $n$-order expansions for $n=2,3, \ldots, 9.$ 
Figure~\ref{fig:marginal} shows that, in most datasets, the performances tend to increase significantly from $n=2$ to $n=3$, but marginally for $n \geq 3$. 
However, somewhat unexpectedly, we also find that some datasets experience 
a small jump (not as high as that from $n=2$ to $n=3$) from $n=7$ to $n=8$ (NDC-classes, coauth-MAG-History) or from $n=8$ to $n=9$ (NDC-substances, coauth-MAG-Geology, coauth-DBLP). 
We speculate that it is because knowing 8 or 9-way interactions 
is often more useful compared to knowing those between 4 to 7-way, for predicting size 10. 
For illustration, papers with 10 authors would be often made by a group of 9 existing collaborators' invitation of another author.

\begin{figure*}[htb]
\vspace{-2mm}
  \centering
  \includegraphics[width=\linewidth]{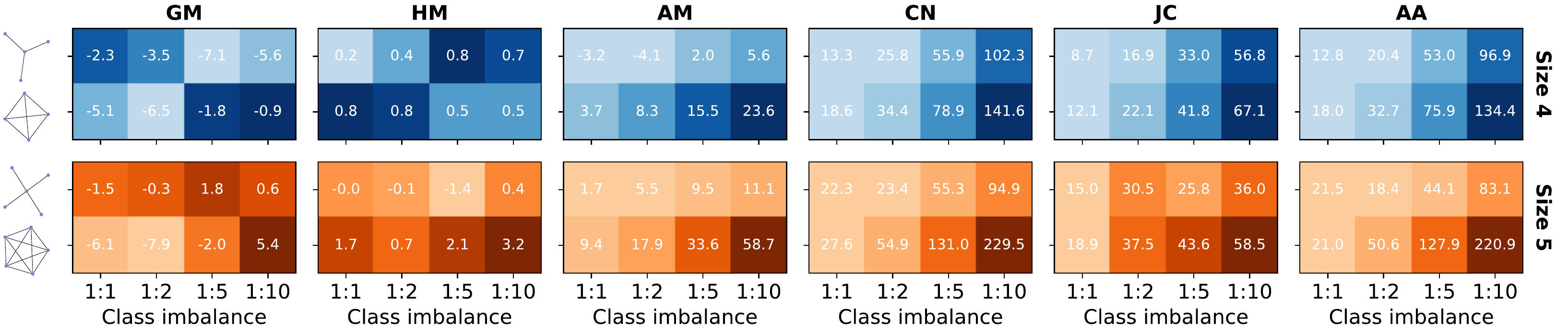}
  \ourcaption{How task difficulty affects the utility of higher-order information. The value in each box is 2 to 3-order gain (\%) averaged across datasets. Boxes with larger gain are colored with greater intensity. We see that as task gets harder, higher-order information helps more.}
\label{fig:heatmap}
\vspace{-0.3cm}
\end{figure*} 

\raggedbottom

\medskip
\vspace{-0.15cm}
\separator
\vspace{-0.15cm}
\noindent{\bf (M2) More hardness of the task gives higher values to higher-order information.}
\vspace{-0.3cm}
\separator
\vspace{-0.1cm}

As discussed in Section ~\ref{subsec:candidate}, we adjust the hardness level in hyperedge prediction by varying the negative set $C_n$ in terms of negative hyperedge types (stars and cliques) and class imbalances (1:1, 1:2, 1:5, 1:10). 
We investigate the impact of those variations on the performance gain, summarized in Figure~\ref{fig:heatmap}. The $y$ axis represents the types of negative hyperedges (stars or cliques), extracted from 2-pg, and the $x$ axis represents different class imbalances. 

Regarding the types of negative hyperedges, we see that the gains from $n=2$ to $n=3$ are larger for cliques than stars. 
As explained in Section~\ref{subsec:candidate}, distinguishing whether a clique is a true hyperedge or not is a much harder task compared to a star.
The troubleshooter is 3-pg, that is, to refer to higher-order information. 
On the impact of class imbalance, again the gain from $n=2$ to $n=3$ is larger, as class imbalance grows. More negative hyperedges imply the increasing hardness in 
obtaining better precision, while maintaining the same sensitivity, and there are more negative hyperedges that resemble positive hyperedges. In such cases, incorporating 3-pg in addition to 2-pg, provided that it gives more information, helps distinguish fake samples better. 
Note that in GM, there are reversed tendencies. This is explained by the property of GM (i.e., geometric mean defined in Section~\ref{subsec:model}) that a pairwise disconnection makes $x_n(c)=0$, i.e., strict stars are easily filtered with only 2-pg, since $x_2(c)=0$.

\vspace{-0.15cm}
\separator
\vspace{-0.15cm}
\noindent{\bf (M3) Higher-order information helps more, when (i) higher-order interactions are more frequent, and (ii) higher-order interactions share less information with pairwise ones.}
\vspace{-0.3cm}
\separator
\vspace{-0.1cm}

\begin{figure}[t!]
\vspace{-2mm}
\centering
\begin{subfigure}{\linewidth}
  \includegraphics[width=\linewidth]{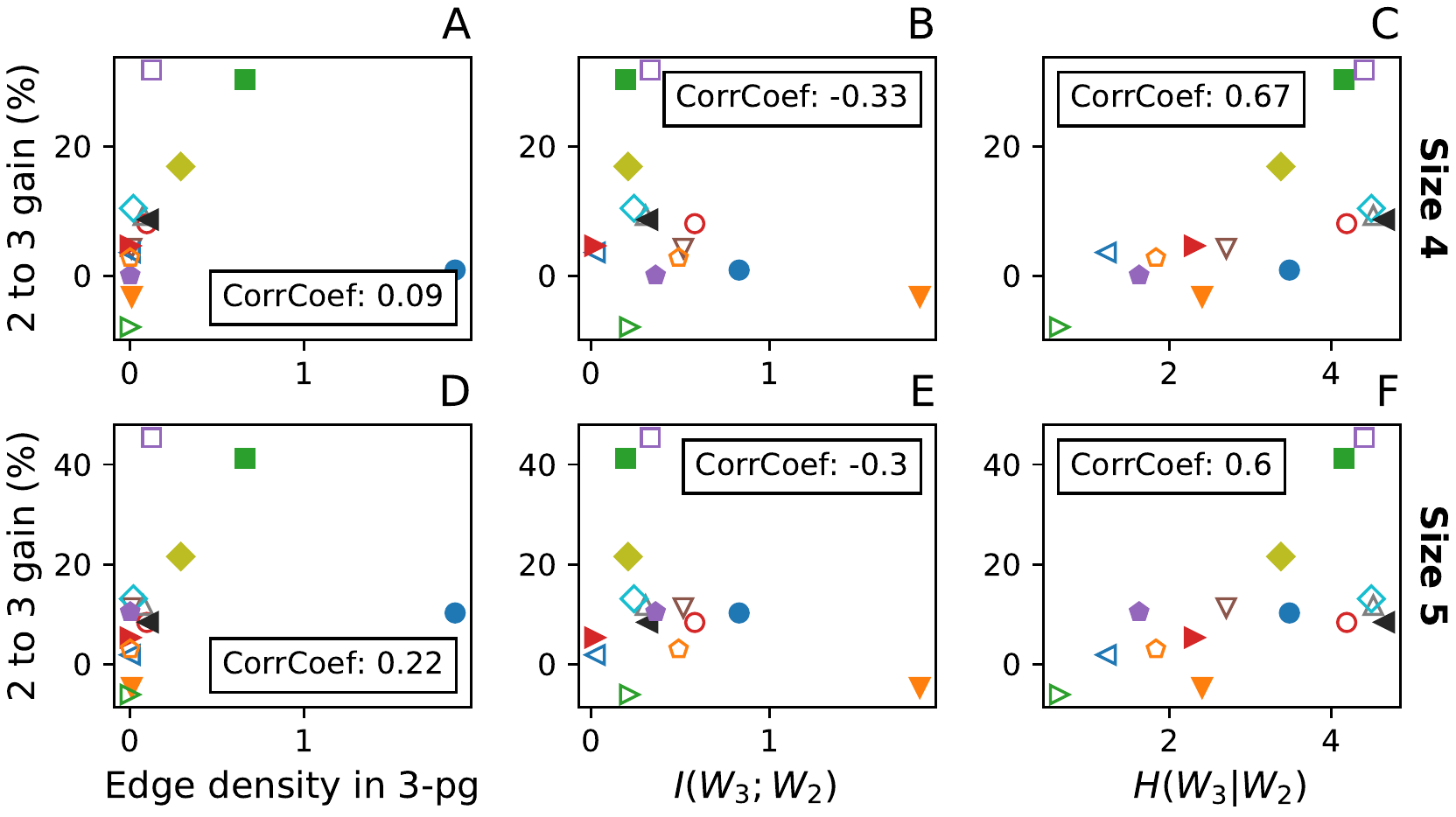}
  \label{fig:clique} 
\end{subfigure}\\ \vspace*{-0.3cm}
\begin{subfigure}{\linewidth}
  \includegraphics[width=1\linewidth]{figures/legend/gn_legend_col3.pdf}
  \label{fig:legend}
\end{subfigure} \\ \vspace*{-0.3cm}
\ourcaption{Interpreting the differences of higher-order gain across datasets. Edge density in 3-pg (A, D) and conditional entropy (C, F) has positive correlation with performance gain. Mutual information (B, E) has negative correlation. Plots are shown for cliques with class imbalance of 1:1. Results on all cases are in \cite{appendix}.}
\label{fig:interpretation}
\end{figure}

We observe different performance gains across different datasets in the two prior messages. 
For example, in Table~\ref{tab:performance}, the average gain from $n=2$ to $n=3$ was about 284\% in contact-primary-school, while it was merely 6.5\% in NDC-classes. We now delve into why they do, for which  
we measure various statistics, including the edge density and information-theoretic values of some pgs.  

\smallskip
\noindent{\em (a) Edge density in 3-pg.}
We first examine edge density in 3-pg, i.e., $\frac{\text{\# edges in 3-pg}}{\text{\# all possible 3-way interactions}}\times100\%$. 
This intuitively quantifies the abundance of 3-way interactions.
We measure the Pearson correlation coefficient between edge density and performance gain, as shown in Figures~\ref{fig:interpretation}-A and D. We observe positive correlations between those two, 0.22 and 0.09 for size 5 and size 4 predictions, respectively, implying that more frequent higher-order interactions let higher-order representation lead to better prediction.

\smallskip
\noindent{\em (b) Mutual information and conditional entropy.}
We next study the aforementioned observation with information-theoretic measures. We expect that adding 3-pg would have larger returns when it contains more information exclusive to itself. 
We set two joint random variables $W_2$ and $W_3$ 
generated by three different nodes sampled $v_1,v_2,v_3 \in V$ uniformly at random, representing a vector of the weights of three pairwise edges and the weight of a triadic edge, i.e.,  
$W_2:=\big(\omega_2(v_1, v_2), \omega_2(v_2,v_3), \omega_2(v_1,v_3) \big),$ and $W_3:=\omega_3(v_1, v_2, v_3).$ 
We consider mutual information $I(W_3; W_2)$ and conditional entropy $H(W_3 \vert W_2)$. Note that $I(W_3; W_2)$ quantifies the amount of shared information between $W_2$ and $W_3$, while $H(W_3 \vert W_2)$ is the remaining information (i.e., uncertainty) of $W_3$ given information of $W_2$.\footnote{ 
Direct estimations of these two measures require a large number of 
samples when the domain spaces of the random variables are large.
We simplify the domain spaces of $W_2$ and $W_3$ by binning them using the function $h(\omega) = \min(\ceil{\log_2{(\omega + 1)}},  9).$
} 

Figures~\ref{fig:interpretation}-B and E show negative correlations -0.33 and -0.3 between the mutual information and performance gain, and similarly, Figures~\ref{fig:interpretation}-C and F show positive correlations 0.67 and 0.6 between the conditional entropy and performance gain. These results imply that 
the gain from $n=2$ to $n=3$ is large when 3-pg is difficult to be explained in terms of 2-pg. We find that the conditional entropy has greater absolute correlations compared to the mutual information. A possible  explanation is that the conditional entropy directly quantifies the information gain while the mutual information focuses on the shared information of 2-pg and 3-pg.

\section{Discussion and conclusion}

In this paper, we studied how much abstraction of group interactions is needed to accurately represent a hypergraph, with hyperedge prediction as our downstream task. We devise the $n$-projected graph to capture the $n$-way interactions in a hypergraph, and express the hypergraph with a collection of $n$-projected graphs. We investigate the performance gain as $n$ grows. We conclude that small $n$ is sufficient due to the diminishing returns, and higher $n$ acts as a troubleshooter in difficult task settings. We provide interpretations why different datasets have different gains. In summary, we investigate 1) how much, 2) when, and 3) why higher-order representations provide better accuracy. We expect that our results would offer insights to relevant works that follow. We leave our source code at \cite{appendix}.

\begin{acks}
This work was supported by the Institute of Information \& Communications Technology Planning \& Evaluation (IITP) grant funded by the Korea government (MSIT) (No.2016-0-00160, Versatile Network System Architecture for Multi-dimensional Diversity).
\end{acks}

\clearpage
\balance
\bibliographystyle{ACM-Reference-Format}
\bibliography{references.bib}


\end{document}